\documentclass[usenatbib]{mn2e}
\usepackage{graphicx,times}
\usepackage{bm,url}
\newcommand{\EQ}{\begin{equation}}
\newcommand{\EN}{\end{equation}}
\newcommand{\EQA}{\begin{eqnarray}}
\newcommand{\ENA}{\end{eqnarray}}
\newcommand{\nab}{\nabla}

\newcommand{\meanB}{\overline{B}}

\newcommand{\meanEMF}{\overline{\mbox{\boldmath ${\mathcal E}$}} {}}

\newcommand{\meanBB}{\bm{\overline{{B}}}}

\newcommand{\meanJJ}{\bm{\overline{{J}}}}
\newcommand{\meanUU}{\bm{\overline{{U}}}}
\newcommand{\meanbb}{\bm{\overline{{b}}}}
\newcommand{\meanuu}{\bm{\overline{{u}}}}

\newcommand{\uu}{\bm{{{u}}}}
\newcommand{\bb}{\bm{{{b}}}}
\newcommand{\jj}{\bm{{{j}}}}

\newcommand{\FF}{\bm{{{F}}}}

\newcommand{\BB}{\bm{{{B}}}}
\newcommand{\UU}{\bm{{{U}}}}

\def\Rm{R_\mathrm{m}}

\newcommand{\Ral}{R_\alpha}
\newcommand{\Qal}{Q_\alpha}
\newcommand{\Rw}{R_\omega}

\newcommand{\etat}{\eta_\mathrm{t}}

\title[Galactic dynamos with stochastic alpha]%
{Galactic dynamo action in presence of stochastic alpha and shear}
\author[S.~Sur, K.~Subramanian]%
{Sharanya Sur$^{1}$, Kandaswamy Subramanian$^{1}$
\thanks{E-mail: sur@iucaa.ernet.in (SS); kandu@iucaa.ernet.in (KS)}\\
$^{1}$Inter-University Centre for Astronomy and
        Astrophysics,  Post Bag 4, Ganeshkhind, Pune 411 007, India
}

\date{}
\begin{document}

\pagerange{\pageref{firstpage}--\pageref{lastpage}} \pubyear{2008}

\maketitle
 
%---------------------------------------------------------------------
\begin{abstract}
Using a one-dimensional $\alpha\omega$-dynamo model appropriate
to galaxies, we study the possibility of dynamo action 
driven by a stochastic alpha effect and shear.
To determine the field evolution, one needs to examine 
a large number of different realizations of the stochastic component
of $\alpha$. The net growth or decay of the field
depends not only on the dynamo parameters but also on the particular realization,
the correlation time of the stochastic $\alpha$ compared to turbulent diffusion
timescale and the time over which the system is evolved. 
For dynamos where both a coherent and fluctuating $\alpha$
are present, the stochasticity of $\alpha$ can help alleviate
catastrophic dynamo quenching, even in the absence
of helicity fluxes. One can obtain final field
strengths up to a fraction $\sim 0.01$ of the equipartition field 
$ B_{eq}$ for dynamo numbers $\vert D\vert \sim 40$, while fields 
comparable to $ B_{eq}$ require much larger degree of $\alpha$ fluctuations
or shear. This type of dynamo may be particularly useful for
amplifying fields in the central regions of disk galaxies.
\end{abstract}
\label{firstpage}

%----------------------------------------------------------------------------------
\section{Introduction}

Large-scale magnetic fields in stars and galaxies are thought to be
generated and maintained by a mean-field turbulent dynamo \citep{Mof78,KR80}.
The potential driver of such mean-field dynamos is the $\alpha$-effect, 
arising whenever one has 
rotation and stratification in a turbulent flow.
Mean-field dynamo (MFD) models using a coherent $\alpha$-effect and
shear have been invoked to explain large-scale fields
observed in disk galaxies \citep{RSS88}.

The possibility of efficient dynamo action arising from random fluctuations in
the $\alpha$-effect in combination with shear was first pointed
out by \citet{VB97}. They investigated a reduced mean-field
dynamo model appropriate to accretion disks and showed 
that growth can occur for 
large enough random fluctuations in alpha.
Several authors have since elaborated various aspects of this stochastic alpha-shear
dynamos \citep{So97,S00,FBR06,Proc07,KR08}.
In particular, \citet{So97} examined a model of a disk dynamo with
a fluctuating alpha antisymmetric in space but which changes sign randomly
with equal probability. 
He argued that intermittent large-scale magnetic fields can grow. 
The role of a stochastic $\alpha$ has also been analyzed in the context of solar
dynamos \citep{Proc07, BS08, Moss08}.

The exact origin of such an incoherent $\alpha$-effect is as yet unclear.
In any large Reynolds number system, 
many degrees of freedom exist, and hence there
could always be a stochastic component of the mean turbulent electromotive
force (emf). This could lead to additive or a multiplicative noise in
the MFD equations. Additive noise provides a seed field for the dynamo,
whereas multiplicative noise in say the $\alpha$ effect, combined with shear,
can lead to exponential growth of the mean field. 
In the solar context, \citet{H93} argued for 
$\alpha$ fluctuations
$\sim u_{0}/\sqrt{M}$, where $u_0$ is the turbulent velocity
and $M$ is the number of cells being averaged over in defining the mean field.
In principle this can be larger than any coherent $\alpha$-effect. 
Multiplicative noise is also seen in simulations which measure
the $\alpha$-effect both in the kinematic regime \citep{SBS08} and in the
nonlinear regime \citep{CH06,BRRK08} and also in direct simulations
of the galactic dynamo \citep{G08}. In fact \citet{BRRK08} measure an incoherent
$\alpha$-effect, with a Gaussian probability density function (PDF), 
even in turbulence driven with a non-helical
forcing, where one does not expect a coherent $\alpha$-effect. 
Combined with shear, such systems show large-scale dynamo action
\citep{BRRK08, Y08}.
Here, we simply examine, in the context of galactic dynamos,
the consequence of having an incoherent alpha effect,  
without considering in detail its exact origin.
The growth of the mean field varies significantly
from one realization of the stochastic process to another,
as also pointed out in \citet{So97}.
It is therefore necessary to examine a large number of realizations of the
stochastic $\alpha$ to determine the efficiency of 
the stochastic $\alpha\omega$-galactic dynamo.  

We outline in section 2, the basics of a 
one-dimensional stochastic $\alpha\omega$-dynamo model appropriate to
galaxies. We present numerical solutions of the
above model in section 3, with two different PDF's for the stochastic alpha;
the first as considered in \citet{So97}
and the second where the stochastic alpha has
a gaussian PDF.
In Section 4, we explore the possibility of alleviating catastrophic 
$\alpha$-quenching in absence of helicity fluxes by including the effects
of a stochastic $\alpha$. Section 5 summarizes our results and the implications
of a stochastic $\alpha\omega$-dynamo for galaxies.  

%--------------------------------------------------------------------------
\section{The stochastic alpha-shear dynamo}

In MFD theory, one starts by splitting the relevant physical
quantities into mean and fluctuating parts, for example 
$\BB = \meanBB + \bb$ for the magnetic field and 
$\UU = \meanUU + \uu$ for the velocity field. The overbars 
denote a suitable averaging procedure with $\meanbb=\meanuu=0$. This results
in the standard mean field dynamo equation
\EQ
\label{indmeanB}
{\partial{\meanBB}\over\partial t}=
\nab\times(\meanUU\times\meanBB+\meanEMF - \eta\nab \times \meanBB), \quad
\nab\cdot\meanBB=0.
\EN
The averaged equation now has a new term,
the mean electromotive force (emf) $\meanEMF={\overline {\uu\times\bb}}$, 
which crucially depends on the statistical properties of the small-scale
velocity and magnetic fields, $\uu$ and $\bb$, respectively.
$\meanUU$ is the mean fluid velocity. Assuming that $\meanBB$
is spatially smooth, $\meanEMF$ can be expressed in terms
of $\meanBB$ and its derivative,
\EQ
\label{meanemf1}
\meanEMF\equiv\overline{\uu\times\bb} = \alpha\meanBB - \eta_{\rm t}\meanJJ
\EN
Here $\meanJJ=\nabla\times\meanBB/\mu_{0}$ (we assume $\mu_{0}=1$ hereafter) 
and $\alpha$ and $\etat$ are turbulent transport coefficients 
that can be expressed in terms of the statistical properties of 
the flow. In the kinematic regime, and assuming isotropic turbulence,
one has 
$\alpha = \alpha_\mathrm{k} = -\frac{1}{3}\tau\overline{\uu\cdot\nabla\times\uu}$, 
and the turbulent diffusion coefficient 
$\etat =\frac{1}{3} \tau\overline{\uu^2}$.
Here, $\tau$, the correlation time of the turbulent velocity
$\uu$, is assumed to be short. 

Since galactic disks are thin, it often suffices to consider a one dimensional
model, where only $z$ derivatives of 
physical variables are retained \citep{RSS88}. 
For the stochastic dynamo which we examine here, 
we also modify the $\alpha$-effect to be of the form:
$\alpha= \alpha_{\rm k} = 
\overline{\alpha}(z) + \alpha_1(z,t)$.
Here $\overline{\alpha}(z)$ is the average $\alpha_{\rm k}$, while
$\alpha_1(z,t)$ is the stochastic $\alpha$ term.  
Therefore, the total $\alpha$  
is a sum of the standard kinetic alpha $\overline{\alpha}$
and a stochastic component $\alpha_1$. 
Further, we consider a mean flow consisting of only a differential rotation such
that ${\meanUU} = (0, r\Omega(r), 0)$.
Then, going to dimensionless variables, Eqn.~(\ref{indmeanB}) 
gives evolution equations for the azimuthal
($\meanB_\phi$) and radial ($\meanB_r$) fields, 
(see also \citet{VB97})
\EQ
\label{ndBr1}
\frac{\partial\meanB_r}{\partial t}=-\frac{\partial}{\partial z}
\left(\Ral g(z) \meanB_\phi + \Qal f(z) N \ \meanB_\phi\right)
+\frac{\partial^2\meanB_r}{\partial z^2},
\EN
\EQ
\label{ndBphi1}
\frac{\partial\meanB_\phi}{\partial t}=
{\Rw\meanB_r} 
+\frac{\partial}{\partial z}
\left(\Ral g(z) \meanB_r +\Qal f(z) N \ \meanB_r\right)
+\frac{\partial^2\meanB_\phi}{\partial z^2}.
\EN
Here the length and time units are $h$ and $t_d = h^2/\etat$
respectively, with $h$ the semi-thickness of the disk.
We adopt $\overline{\alpha} = \alpha_0 g(z) $, and 
$\alpha_1=\alpha_s f(z) N(t) $ where 
$f(z) = g(z) = \sin(\pi z)$ 
takes care of the symmetry condition. 
$N(t)$ is a stochastic function.
In our numerical solutions we adopt the following
procedure: We split $t$ into equally spaced intervals $[n\tau_c,(n+1)\tau_c]$,
where $\tau_c$ is the correlation time of the stochastic alpha,
and $n=0,1,2....$ are integers.
And in any such time interval $N$ is a random number chosen from a Gaussian (or some other) 
probability distribution, with unit variance. 
The relevant dynamo control parameters are $\Ral, \Qal$ and $\Rw$ defined as 
\EQ
\Ral=\frac{{\alpha_0}h}{\etat}\,,\hfil
\Qal=\frac{{\alpha_s}h}{\etat}\,,\hfil
\Rw=\frac{Gh^2}{\etat}\,,\hfil
\EN
Here $G = rd\Omega/dr = -\Omega$, for a flat rotation curve.
From Krause's formula, $\alpha_0\simeq l^{2}_{0}\Omega/h$,
where $l_0$ is the integral scale of interstellar turbulence.
Then $\Ral \sim 3 \Omega\,t_{ed}$, assuming $\etat \sim l_0u_0/3$
and $\tau\sim t_{ed}=l_0/u_0$, the eddy turnover time. 
Typical values of the dynamo control parameters in 
the solar neighborhood of the Milky Way could be 
$\Ral \sim 1.0$, and $|\Rw| \sim 10 - 15$, corresponding
to a "dynamo number" $D = \Ral\Rw \sim -10$ to $D \sim -15$ \citep{RSS88}. 
The strength of $\Qal \sim 3 (h/l_0)M^{-1/2}$ if one uses the estimate of \citet{H93}.
A horizontal average over a scale $h$ to define $\meanBB$ 
(cf. \citet{BRRK08}) would suggest 
$M \sim (h/l_0)^2$ and hence $\Qal \sim 3$.
However since the exact origin of such fluctuations is as yet unclear,
we will vary $\Qal$ around these values.
Thus in general, we will have $|\Rw|\gg\Ral,\Qal$ so that one can make
the standard $\alpha\omega$-dynamo approximation, where
one neglects the terms with co-efficients $\Ral$ and $\Qal$ in
Eqn.~(\ref{ndBphi1}). 

Note that $\Omega\propto1/r$, and thus one can have larger
dynamo parameters towards the disk centre,
depending also on how $h$ and $l_0$ behave there.
The disk height could be smaller, but $l_0$ could also be smaller
in the denser inner galactic regions, where supernovae are more
confined. This could lead to 
a net increase in $\Rw \propto \Omega h^2/l_0$.
Any increase in $\Ral$ depends on how much $l_0$ decreases
compared to the increase in $\Omega$. Changes in $\Qal$
depend on the origin of the $\alpha$ fluctuations.
For example, if $h$ decreases by factor $2$ and $l_0$ decreases
a factor $5$ in the inner galaxy, $\Rw$ would increase by
a factor $6.25 (r/2 {\rm kpc})^{-1}$ and $\Ral$ or $\Qal$ would remain
about the same, compared to the solar neighborhood.
Overall larger dynamo numbers can 
be expected in the inner regions of disk galaxies.
We now turn to the solution of the stochastic $\alpha\omega$-
dynamo equations.

%%%%%%%%%%%%%%%%%%%%%%%%%%%%%%%%%%%%%%%%%%%%%%%%%%%%%%%%%%%%%%%%%%%%%%%%
\begin{figure}\begin{center}
\includegraphics[width=\columnwidth]{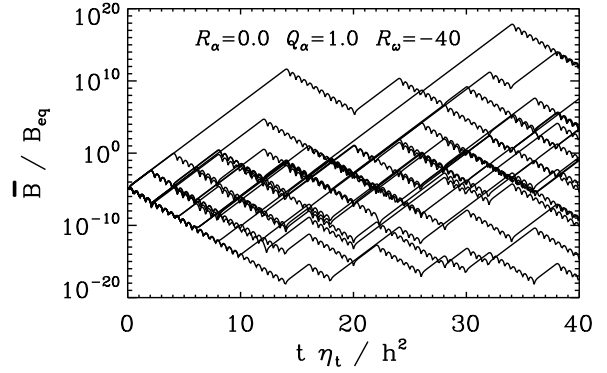}
\end{center}\caption[]{
Time evolution of the rms mean magnetic field in the Sokoloff $1997$
model for different realizations using a long correlation time for 
$\alpha_1(t)$. Parameter values used are $\Ral=0.0$,
$\Qal=1.0$, and $\Rw=-40$. 
}\label{dmqalrw}\end{figure}
%%%%%%%%%%%%%%%%%%%%%%%%%%%%%%%%%%%%%%%%%%%%%%%%%%%%%%%%%%%%%%%%%%%%%%%%%

%%%%%%%%%%%%%%%%%%%%%%%%%%%%%%%%%%%%%%%%%%%%%%%%%%%%%%%%%%%%%%%%%%%%%%%%
\begin{figure}\begin{center}
\includegraphics[width=\columnwidth]{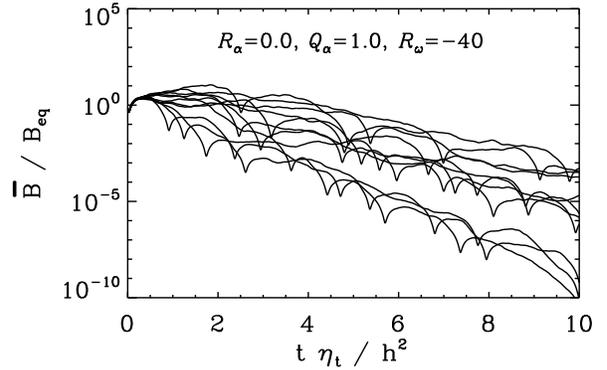}
\end{center}\caption[]{
Time evolution of the rms mean magnetic field in different
realizations using a short correlation time for $\alpha_1(t)$ 
and a gaussian PDF. Plots are obtained for parameter values $\Ral=0.0$, 
$\Qal=1.0$, and $\Rw=-40$. 
}\label{modqalrw}\end{figure}
%%%%%%%%%%%%%%%%%%%%%%%%%%%%%%%%%%%%%%%%%%%%%%%%%%%%%%%%%%%%%%%%%%%%%%%%%

%%%%%%%%%%%%%%%%%%%%%%%%%%%%%%%%%%%%%%%%%%%%%%%%%%%%%%%%%%%%%%%%%%%%%%%%
\begin{figure}\begin{center}
\includegraphics[width=\columnwidth,bbllx=18bp,bblly=165bp,bburx=585bp
,bbury=706bp]{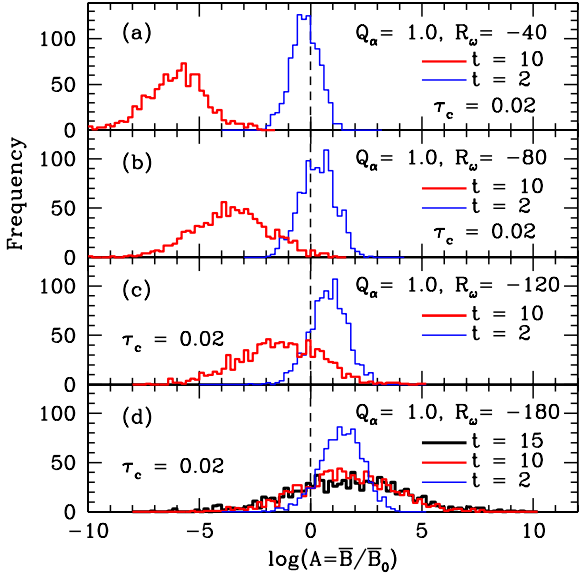}
\end{center}\caption[]{
Frequency distribution of the dynamo amplification 
$A= \meanB/\meanB_0$ at $t=2$ (blue/thin), $t=10$ (red/normal) and also
at $t=15$ (black/thick) (in panel d), for $1000$ realizations of $\alpha_1(t)$
in each histogram. Here 
$\Ral=0.0$, $\Qal=1.0$, and $\Rw=-40, -80, -120$ 
and $-180$. Bin size is $0.2$.
}\label{histo}\end{figure}
%%%%%%%%%%%%%%%%%%%%%%%%%%%%%%%%%%%%%%%%%%%%%%%%%%%%%%%%%%%%%%%%%%%%%%%%%

\section{Numerical Solutions}
Our primary interest is in a scenario where large-scale dynamo action is
possible in presence of stochastic alpha and shear. Thus we first seek
numerical solutions to Eqs.~(\ref{ndBr1}) and (\ref{ndBphi1}), 
in the $\alpha\omega$-dynamo approximation, with the coherent part of
the $\alpha$-effect taken to be zero; that is with $\Ral = 0$.
The code uses 
a 6th order explicit
finite difference scheme for the space-derivatives and 
3rd order accurate time-stepping
scheme; see 
\citet{B03} for details. 
We use vacuum boundary conditions for the fields
\EQ\label{vacond}
\meanB_r=\meanB_\phi=0 \quad\mbox{at $z=\pm h$}
\EN

As a test case, we numerically implemented
the \citet{So97} model with $\bar\alpha=0$ and
$\alpha_1=\alpha_s f(z)N(t)$; $N$ being either
$+1$ or $-1$ with equal probability in any time interval $n\tau_c < t < (n+1)\tau_c$. 
For $N=1$, the system 
behaves as a standard $\alpha\omega$-dynamo
with growing solutions, while for $N=-1$ we have decaying oscillations.
So if the system is 
evolved over a finite time interval, there would
be random instances of growth and decay. If $\gamma$ is the growth rate of the
growing solutions and $-\,\zeta$ that of the decaying ones, the ensemble averaged
growth rate is $\Gamma=(\gamma-\zeta)/2$  \citep{So97}. 
Thus when $\gamma>\zeta$, one obtains $\Gamma>0$ resulting in an overall growth
above a critical dynamo number $D_c$. To estimate $D_c$,
we use the perturbation solutions discussed in \citet{SSS07}. This gives 
$\gamma \approx -\pi^2/4 + \sqrt{\pi|D|}/2$ and $\zeta\approx\pi^2/4$,
and thus 
$D_{c}\approx-\pi^3$ for the \citet{So97} model.
This is somewhat larger in magnitude than
the critical dynamo number $\sim \pi^3/4$, which obtains for the coherent 
$\alpha\omega$ dynamo (by demanding $\gamma > 0$).

These features are illustrated in Fig.~(\ref{dmqalrw}).
Here we have chosen $\tau_c=2t_d$ 
so that one can clearly see both the growing and
decaying phases and their net effect.
Starting with random seed fields $\meanB_r, \meanB_\phi \sim 10^{-6}$, 
we find a number of growing as well as decaying realizations
for moderate dynamo number $D = -40$. 
It is evident from Fig.~(\ref{dmqalrw}), that any given realization has periods,
$N_+\tau_c$, of steady growth (when $N=1$) and periods, $N_-\tau_c$ 
of oscillatory decay (when $N=-1$).
One gets a net growth of the field in about $65\%$ of the realizations, 
as roughly expected from the above arguments for $\vert D\vert > \vert D_c\vert$.
For a larger magnitude of the dynamo number $\vert D\vert$ one gets
a greater probability for growth.
We have also examined the opposite limit when $\tau_c < t_d$, and find that
the dynamo becomes less efficient (see below).
These solutions clearly demonstrate the basic idea behind the
incoherent $\alpha\omega$ dynamo as discussed in \citet{So97},
that one needs to consider many realizations of the stochastic process. 
Just solving a double averaged version of the MFD equations need
not be representative of the actual evolution of the dynamo for a given
realization.

Of course the PDF of the stochastic alpha is not expected to be as described
in \citet{So97}; for example \citet{BRRK08} found it can be approximated
as a Gaussian. Also in general we expect $\tau_c < t_d$.
We now present the results obtained
by solving Eqs.~(\ref{ndBr1}) and (\ref{ndBphi1}) with the random number $N$ for
$\alpha_1$ chosen from a gaussian PDF, and
adopting $\tau_c = 0.02 t_d$. 
This is about $1.5t_{ed}$, 
assuming $h = 500$ pc, $\etat = 10^{26}$ cm$^2$ s$^{-1}$, $l_0 = 100$ pc, and
$u_0 \sim 10$ km s$^{-1}$. The initial seed fields are random with
amplitudes of $O(1)$. 
The MFD equations were evolved up to 10 turbulent diffusion time scales,
$t=10$, for dynamo numbers $D=-40, -80$ and $-120$ and upto $t=15$ for 
$D=-180$. We also considered 1000 realizations of $\alpha_1(t)$ for each
$D$ so as to obtain good statistics. 
Note that to probe the PDF of the 
dynamo amplification upto a $3\sigma$ level
one needs about these many realizations.

Fig.~(\ref{modqalrw}) shows the time evolution of the RMS (large scale)
magnetic field, $\meanB$, for a subset of realizations with 
$\Ral=0.0, \Qal=1.0$ and $\Rw=-40$. There is an initial
decay of $\meanB$, while the system discovers
the proper eigenfunction. 
Further evolution then occured on the diffusion
time-scale $t_d$. 
In all realizations, $\meanB$ shows an oscillatory decay, even though 
a significant number of realizations showed growth of $\meanB$ 
up to $t=2$. For higher dynamo numbers, growth is sustained 
for a longer time and for a larger number of realizations. 
We find that the \citet{So97} model also
shows similar features for short correlation times.
Thus having $\tau_c \ll t_d$,
qualitatively changes the behavior of the
dynamo and leaves an imprint of $t_d$ in the system
evolution rather than $\tau_c$.
In order to have a quantitative measure of how many realizations
show net growth, we show in Fig.~ \ref{histo} the frequency distribution of
the dynamo amplification $A= \meanB/\meanB_0$ at $t=2$, $t=10$ 
and also at $t=15$ in panel (d) 
for 1000 realizations of $\alpha_1(t)$, at dynamo numbers,
$D=-40, -80, -120$ and $-180$. Here $\meanB_0 =0.32 - 0.35$ is roughly the value
to which $\meanB$ initially decays in all the realizations.
For $D=-40, -80$ and $-120$, we obtain 
$A > 1$, for respectively $34\%, 65\%$ and $82.8\%$ of realizations at
$t=2$. However at a later time $t=10$, this percentage decreases to 
$0\%, 18\%$ and $24\%$ respectively.
This is evident in the gradual shift of the histogram to the left.
For $\vert D\vert= 160 - 180$ the PDF of $\vert A\vert$ remains 
stationary at late times; see panel (d). Above this
range, the mean amplification secularly increases with time. 
Thus, our results show that a stochastic $\alpha\omega$-dynamo is reasonably 
efficient over a few $t_d$ even at $\Rw =-40$, but requires 
much larger dynamo numbers, as plausible towards galactic centres, 
to sustain fields for long periods.

%---------------------------------------------------------------------------------------
\section{Dynamical alpha quenching of the stochastic dynamo}

Conservation of magnetic helicity is regarded as a key constraint in
the evolution of large-scale magnetic fields (see \citet{BS05} for a review). 
A consequence of helicity conservation is the production of 
equal and opposite amounts of magnetic helicity 
in $\meanBB$ and $\bb$ by the turbulent emf $\meanEMF$.
Closure models then 
imply a suppression of dynamo action
due to the growing current helicity associated with $\bb$ \citep{PFL76,KR82,BF02}. 
The effect of the small-scale magnetic field on the
total $\alpha$-effect is described by the addition of
a magnetic alpha to the kinetic alpha, 
$\alpha = \alpha_\mathrm{k}+\alpha_\mathrm{m}$. 
Here $\alpha_\mathrm{k}$ represents the kinetic $\alpha$-effect
and $\alpha_\mathrm{m}
= \frac{1}{3}\rho^{-1}\tau\overline{\jj\cdot\bb}$ is the magnetic
contribution to the $\alpha$-effect, with $\rho$ the fluid density.
Specifically, the growth of the magnetic
alpha ($\alpha_{\rm m}$) to cancel the kinetic alpha 
($\alpha_{\rm k}$) results in a suppression of the total $\alpha$-effect.
This suppression can be catastrophic in the sense that the large-scale field is
quenched in an $\Rm$ dependent manner. Helicity fluxes
across the boundaries of the disk have been identified as a possible mechanism
to shed small-scale magnetic helicity, and prevent such quenching 
\citep{BF00,KMRS00,BF01,VC01,B05,SH06,SSS07}. 

This situation could change in the 
presence of a stochastic component, as the kinetic alpha can
undergo frequent sign reversals. Hence by the time the $\alpha_{\rm m}$
grows to cancel $\alpha_\mathrm{k}$, the kinetic alpha itself might have
changed sign. It is then of interest to ask whether addition of a stochastic
component to the kinetic alpha can stem the catastrophic quenching. This would
then naturally provide a mechanism for healthy dynamo action even in the absence of
helicity fluxes. The numerical analysis of the previous section is therefore 
extended by including an $\alpha_{\rm m}$ contribution to $\alpha$ in 
equations (\ref{ndBr1}) and (\ref{ndBphi1}) supplemented with an evolution
equation for $\alpha_{\mathrm m}$. 
This can readily be motivated by considering the
helicity conservation equation written in terms of the helicity density
$\chi$ of the small-scale magnetic field \citep{SB06}, 
\EQ
\label{hconv}
\frac{\partial \chi}{\partial t} +\nabla\cdot\FF = 
-2\meanEMF\cdot\meanBB-2\eta\overline{\jj\cdot\bb}.
\EN
Here $\FF$ is the helicity flux density. 
Retaining only the $z$-derivatives and using the fact that the 
main contribution to $\alpha_{\rm m}$ comes from the integral scale
of turbulence 
\citep{SH06}, so that 
$\overline{\jj\cdot\bb}\simeq k_0^{2}\overline{{\bm a}\cdot\bb}$
\EQ
\label{amchi}
\alpha_{\rm m} \simeq\frac{1}{3}\tau \,{k_0^2} \,\frac{\chi}{\rho}\,.
\EN
Eqn.~(\ref{hconv}) can be expressed in dimensionless form, by measuring
$\alpha$ in units of $\alpha_0$ and the magnetic field in units of
$B_{eq}$, where $B_{eq}^2 = \rho \overline{u^2}$. 
In the absence of helicity fluxes, i.e $\FF=0$ we have, 
\EQ
\label{alpm1}
\frac{\partial\alpha_{\rm m}}{\partial t}=-C\left(\left(g
+\frac{\Qal}{\Ral}fN + \alpha_{\rm m}\right)\meanB^2 
-\frac{\meanJJ\cdot\meanBB}{\Ral}
+\frac{\alpha_{\rm m}}{ \Rm}\right)
\EN
where $\Rm = \eta_t/\eta$, $C=2\pi^2(k_0/k_1)^2$, $k_1=\pi/h$ and we take 
$k_1/k_0=5$. Further, $\meanJJ\cdot\meanBB$ is
the current helicity density of the large-scale field and
is given by
\EQ
\label{JB}
\meanJJ\cdot\meanBB=
                \meanB_\phi{\partial\meanB_r\over\partial z}
                -\meanB_r{\partial\meanB_\phi\over\partial z}.
\EN
We adopt $\alpha_{\rm m}=0$ at $t=0$ and random initial fields
of $O(10^{-6})$. 
The system of equations (\ref{ndBr1}), (\ref{ndBphi1})
and (\ref{alpm1}) are then solved numerically in the $\alpha\omega$-dynamo
approximation. Note that there is an extra term 
$-\partial(\Ral\alpha_{\rm m}\meanB_\phi)/\partial z$
in Eq.~\ref{ndBr1} and no helicity fluxes are added to the r.h.s of
Eqn.~(\ref{alpm1}). 

%%%%%%%%%%%%%%%%%%%%%%%%%%%%%%%%%%%%%%%%%%%%%%%%%%%%%%%%%%%%%%%%%%%%
\begin{figure}\begin{center}
\includegraphics[width=\columnwidth]{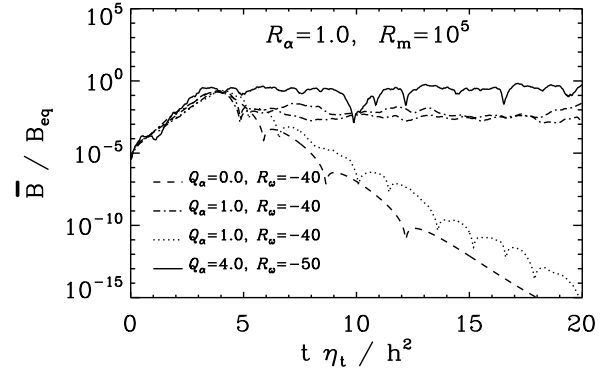}
\end{center}\caption[]{
Time evolution of the rms mean magnetic field for different
realizations in the dynamical $\alpha$-quenching model 
with parameter values $\Ral=1.0, \Qal=0.0-4.0, 
|\Rw|=40-50$ and $\Rm=10^5$. 
}\label{c200ral1qal1rw40}\end{figure}
%%%%%%%%%%%%%%%%%%%%%%%%%%%%%%%%%%%%%%%%%%%%%%%%%%%%%%%%%%%%%%%%%%%%%%

%%%%%%%%%%%%%%%%%%%%%%%%%%%%%%%%%%%%%%%%%%%%%%%%%%%%%%%%%%%%%%%%
\begin{figure}\begin{center}
\includegraphics[width=\columnwidth]{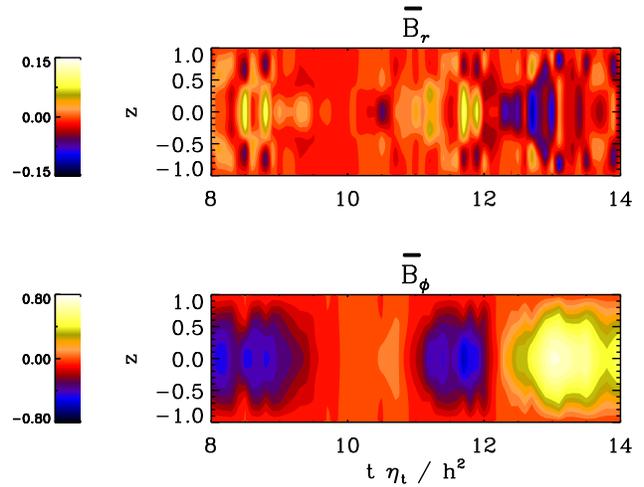}
\end{center}\caption[]{
Space-time diagrams of the radial and azimuthal components of the
large-scale field for a realization with parameter
values $\Ral=1.0, \Qal=4.0$ and $|\Rw|=50$.
The color bars on the left panel shows the magnitude of the field.
}\label{contour}\end{figure}
%%%%%%%%%%%%%%%%%%%%%%%%%%%%%%%%%%%%%%%%%%%%%%%%%%%%%%%%%%%%%%%%

Fig.~(\ref{c200ral1qal1rw40}) shows the time evolution of the RMS large-scale 
field in a number of realizations with $\Ral=1.0, \Qal=0.0-4.0, |\Rw|=40-50$
and $\Rm=10^5$. Note that for $\Qal=0.0$, $\Ral=1.0$ and $\Rw=-40$ (shown by dashed lines),
we recover the standard result that the magnetic field is catastrophically
quenched to very low values. The catastrophic quenching still obtains in
some realizations for a moderate value of $\Qal=1.0$ (shown in dotted line
in the figure). But in other realizations (shown by dash-dotted lines in the
figure), a stochastic kinetic alpha alleviates this quenching to obtain fields
of order $0.01-0.001 B_{eq}$. 
A detailed analysis shows that, for these dynamo parameters,  
$\meanB$ has a net growth in about $13\%$ of all the realizations,
even till $t=20$.
In fact, stronger values of $\Qal$ and $\Rw$ can even 
amplify the field to near equipartition values. Such an example, 
adopting $\Qal=4.0$ and $|\Rw|=50$ is shown by the solid line in the above figure. 
A space-time diagram for this realization, between times $t=8-14$, 
is shown in Fig.~\ref{contour}. Both the radial and azimuthal fields have 
quadrupolar symmetry and show several reversals in sign during this period.
We recall that high values of the dynamo control parameters are plausible 
towards the central regions of a galaxy. Therefore a stochastic
$\alpha\omega$-dynamo is more likely to grow coherent magnetic
fields efficiently towards the central regions of disk galaxies.

%-----------------------------------------------------------------------------------------
\section{Conclusions}
We have examined here how a stochastic $\alpha$-effect in association
with shear can lead to the generation of  large-scale galactic magnetic fields.
To determine the field evolution, one needs to examine a large number of different
realizations of the stochastic $\alpha_1(t)$. 
The net growth or decay of the field depends on the particular realization, 
the correlation time of the stochastic $\alpha$ compared to turbulent diffusion
timescale and the time over which the system is evolved. 

The results are illustrated first with the simple model 
of \citet{So97} in Fig.~\ref{dmqalrw}. 
Here the magnitude of $\alpha_1$, takes randomly a value
$+\alpha_s g(z)$ or $-\alpha_s g(z)$ over any time interval $\tau_c$, 
with equal probability. 
Any given realization of $\alpha_1(t)$ will have $N_+$ periods of steady growth
(when $N=1$) and $N_-$ periods of oscillatory decay (when $N=-1$).
But, since the growth rate ($\gamma$) and decay rates ($\zeta$) are different,
this could lead to a net growth or decay as pointed out by \citet{So97}.
The critical dynamo number for getting growth
for say $50\%$ of realizations is $\vert D_c\vert \sim 30$, moderately
larger than $\vert D_c\vert \sim 10$ required for the coherent $\alpha\omega$-dynamo.
Our numerical solutions confirm the applicability of this picture
for long correlation times $\tau_c=2t_d$, while for
$\tau_c < t_d$, the picture changes qualitatively,
and the dynamo becomes less efficient.

We then examined more realistic MFD models 
with a short correlation time ($\tau_c = 1.5 t_{ed}$), for a stochastic
$\alpha_1(t)$ chosen from a gaussian PDF. 
Our results are given in Fig.~\ref{modqalrw} and Fig.~\ref{histo}.
In this case as well, 
and for a stochastic $\alpha\omega$-dynamo number $D=-40$, 
about $34\%$ of realizations showed 
growth of $\meanB$ till $t \sim 2$. However subsequently $\meanB$ decays
to negligible values by $t \sim 10$.
For higher dynamo numbers, growth is sustained
for a longer time and for a larger number of realizations (cf. Fig.~\ref{histo}).
One requires $ D \sim -120$ to obtain long term growth (till $t \sim 10$),
in a significant number ($\approx$ 24\%) of realizations,
$\vert D \vert \approx 160 - 180$ for 
the PDF of $\vert A\vert$ to remain stationary and a larger $\vert D\vert$ 
for secular growth at late times.
Note that having an additional coherent $\alpha$, with $\vert\Ral\Rw\vert > 10$, 
would ensure growth of $\meanBB$. However, the quenching imposed by helicity
conservation still needs to be alleviated.

In the usual $\alpha\omega$-dynamo, such helicity conservation
leads to a growth of the magnetic $\alpha_{\rm m}$, which 
tends to cancel the kinetic $\alpha_{\rm k}$, so as to
catastrophically quench the dynamo. This problem can be alleviated
by having fluxes of magnetic helicity.
In contrast, for a stochastic $\alpha\omega$-dynamo, by the time
$\alpha_{\rm m}$ has grown, $\alpha_{\rm k}$ could have changed sign.
This raises the possibility of alleviating catastrophic $\alpha$-
quenching without helicity fluxes.
To examine this possibility, we solved the stochastic $\alpha\omega$-dynamo equations 
along with the dynamical $\alpha$-quenching equation. We included both a coherent
and incoherent $\alpha$-effect. 
In general the radial and azimuthal fields are again quadrupolar
and show occasional reversals in time (see Fig. \ref{contour}).
When the coherent and stochastic components of $\alpha$ are comparable, 
and even without any helicity flux,
we find that steady large-scale magnetic field of strengths of about $0.01B_{eq}$
could be obtained in some realizations even for $D \sim - 40$.
Field strengths above $0.3B_{eq}$ are obtained for stronger amplitude
of random $\alpha$-fluctuations in association with strong shear 
(Fig. \ref{c200ral1qal1rw40}).
Therefore, a stochastic $\alpha\omega$-dynamo
model is more likely to grow large-scale magnetic fields efficiently towards
the central regions of a galaxy and even in the absence of helicity fluxes.

We have focussed on the application of a stochastic $\alpha\omega$-dynamo
model to galaxies. 
However, our emphasis on examining a large number of
realizations and our results on alleviating $\alpha$-quenching with a 
random $\alpha$ could be applicable to other astrophysical dynamos as well.
More work is needed to elucidate the origin of the incoherent $\alpha$
and also study the influence of spatial decorrelation on the dynamo.

\section*{Acknowledgments}
We acknowledge Nordita and the KITP for providing a stimulating atmosphere
during their programs on dynamo theory.
This research was supported in part by the National Science
Foundation under grant PHY05-51164. We thank Axel Brandenburg, Eric Blackman,
David Moss, Anvar Shukurov and Dmitri Sokoloff for valuable comments.
SS thanks Council of Scientific and Industrial Research, India for financial
support. 
%------------------------------------------------------------------------------
\newcommand{\ybook}[3]{ #1, {#2} (#3)}
\newcommand{\yjfm}[3]{ #1, {J.\ Fluid Mech.,} {#2}, #3}
\newcommand{\yprl}[3]{ #1, {Phys.\ Rev.\ Lett.,} {#2}, #3}
\newcommand{\ypre}[3]{ #1, {Phys.\ Rev.\ E,} {#2}, #3}
\newcommand{\yapj}[3]{ #1, {ApJ,} {#2}, #3}
\newcommand{\yan}[3]{ #1, {AN,} {#2}, #3}
\newcommand{\yana}[3]{ #1, {A\&A,} {#2}, #3}
\newcommand{\ygafd}[3]{ #1, {Geophys.\ Astrophys.\ Fluid Dyn.,} {#2}, #3}
\newcommand{\ypf}[3]{ #1, {Phys.\ Fluids,} {#2}, #3}
\newcommand{\yproc}[5]{ #1, in {#3}, ed.\ #4 (#5), #2}
\newcommand{\yjour}[4]{ #1, {#2} {#3}, #4.}
\newcommand{\ymn}[3]{ #1, {MNRAS,} {#2}, #3}
\newcommand{\ymnl}[3]{ #1, {MNRAS,} {#2}, #3}
\newcommand{\ysolp}[3]{ #1, {Sol. Phys.,} {#2}, #3}
\newcommand{\sapj}[1]{ #1, {ApJ,} (submitted)}
\newcommand{\asrep}[3]{#1, {Astronomy Reports,} {#2}, #3}

\label{lastpage}
\end{document}